\documentclass[fleqn,10pt]{wlscirep}
\usepackage[utf8]{inputenc}
\usepackage[T1]{fontenc}
\usepackage{amsmath,graphicx,hyperref,xcolor}

\newcommand\figref{Fig.~\ref}

\title{Optimal photon pairs for quantum communication protocols}

\author[1,*]{Miko\l aj Lasota}
\author[1]{Piotr Kolenderski}
\affil[1]{Faculty of Physics, Astronomy and Informatics, Nicolaus Copernicus University, Grudziadzka 5, 87-100 Toru\'{n}, Poland}

\affil[*]{miklas@umk.pl}

\begin{abstract}
We theoretically investigate the problem of finding optimal characteristics of photon pairs, produced in the spontaneous parametric down-conversion (SPDC) process, for fiber-based quantum communication (QC) protocols. By using the accessible setup parameters, the pump pulse duration and the extended phase-matching function width,  we minimize the temporal width of SPDC photons within the general scenario. This allows one to perform more effectively the temporal filtering procedure, which aims at reducing the noise acquired by the measurement devices. Moreover, we compare the obtained results with the achievable parameter values for SPDC sources based on $\beta$-Barium Borate (BBO) crystal. We also investigate the influence of non-zero detection timing jitter. Finally, we apply our optimization strategy to a simple quantum key distribution scheme to show that the full optimization of an SPDC source can potentially extend the maximal security distance  by several tens of kilometres, which is around 30\% more as compared to previous approaches. 
\end{abstract}
\begin{document}

\flushbottom
\maketitle

\thispagestyle{empty}

\section*{Introduction}

Quantum communication is a vast field of physical science focused on improving the process of information distribution among spatially separated entities with the use of quantum mechanics. The exploration of various types of quantum correlation and application of fundamental quantum laws has lead to a plethora of proposals for novel communication protocols, including quantum key distribution \cite{Bennett1984,Ekert1991}, secret sharing \cite{Hillery1999}, quantum teleportation \cite{Bennett1993} and dense coding \cite{Bennett1992c}. However, the initially proposed theoretical versions of these protocols have typically assumed ideal performance of the setup elements required for their physical realization, which is unreachable in practice. As a consequence, the performance of real-life implementations of QC protocols has been severely limited.

Realization of many such schemes requires using sources of single photons or entangled photon pairs. One of the most popular types of them are the devices utilizing the phenomenon of spontaneous parametric down-conversion \cite{Louisell1961,Burnham1970}. They have many advantages, including high quality of the emitted photons \cite{Fasel2004,Bock2016}, high generation and collection efficiency \cite{Pomarico2012,Ramelow2013,Kaneda2016} and relatively low cost of their construction. Therefore, they have been extensively used in practical implementations of many QC protocols \cite{Mattle1996,Bouwmeester1997,Pan1998,Jennewein2000,Tittel2001}. However, photons born in the SPDC process are not monochromatic. Thus, they propagate through dispersive media (\emph{e.g.} the standard telecommunication fibers) with wavelength-dependent velocity. As a consequence, their temporal width, defined as the standard deviation of the probability distribution function for the time of their arrival at the destination point, grows with the length of the utilized dispersive quantum channel. It forces the experimenter to define longer detection windows for the photon measurement system in order not to lose considerable amount of real signals. However, the longer detection windows are, the more noise is registered during the realization of a given QC protocol, negatively affecting its performance.

Taking into account the above consideration, the minimization of the temporal width of the emitted photons after their propagation through telecommunication fibers is very important. Nevertheless, to the best of our knowledge such a general  optimization has not yet been done. Only recently it was shown that changing the properties of pairs of photons can indeed significantly influence the performance of quantum protocols \cite{Sedziak2017,Lasota2018,Sedziak2019}. A preliminary optimization of SPDC source for a specific setup configuration was also performed\cite{Sedziak2019}. In this manuscript we generalize this optimization to an arbitrary QC scheme. We also investigate the influence of non-zero timing jitter of realistic detectors on the obtained results and discuss whether the theoretically optimal values can be implemented in practice. Finally, we estimate the advantage stemming from the optimization of SPDC source on the maximal security distance of a basic quantum key distribution (QKD) scheme.

\begin{figure}[t]
	\centering
	\begin{minipage}{0.45\textwidth}
			\centering
			\includegraphics[width=0.98\columnwidth]{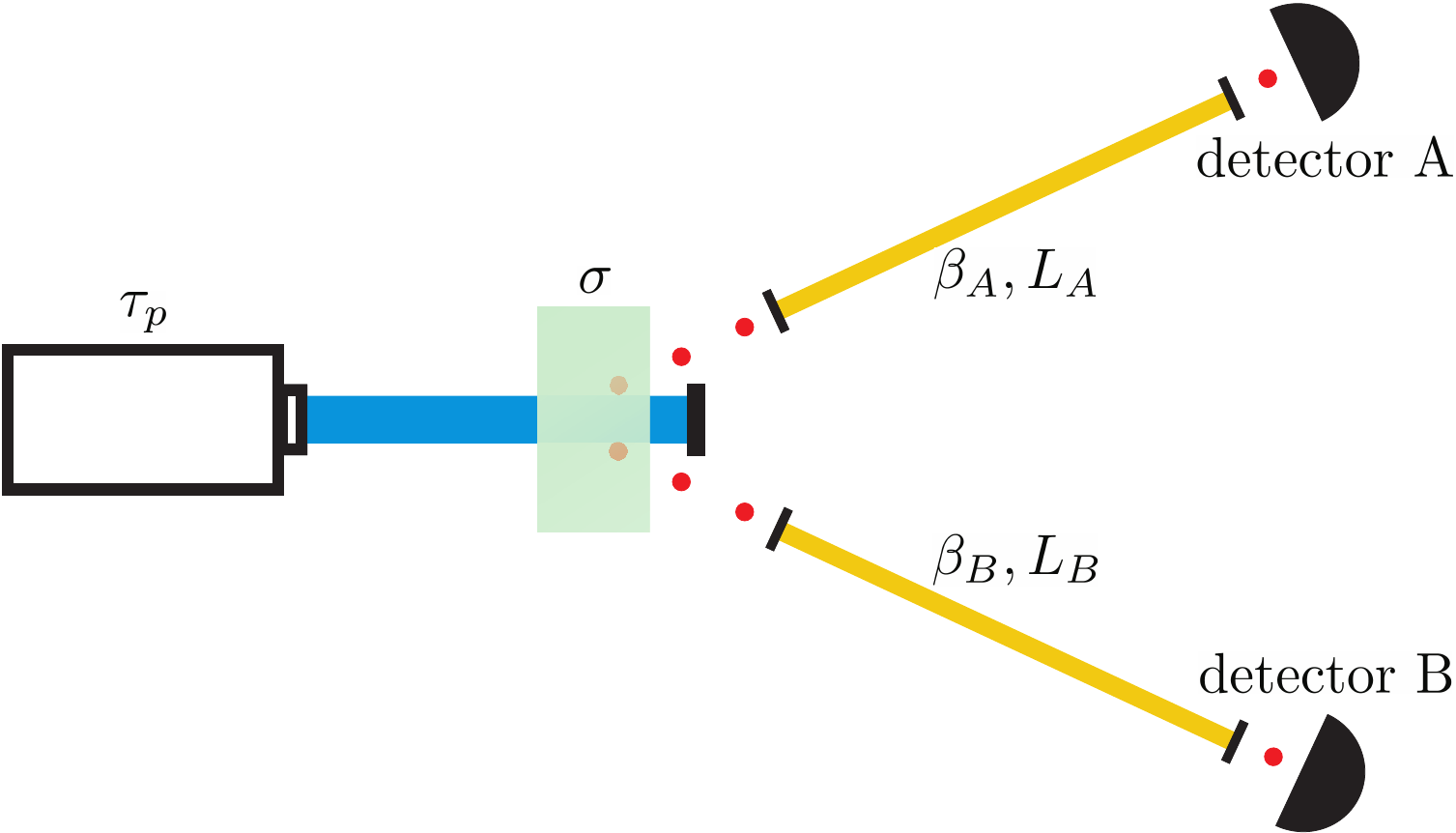}
			\caption{{\bf Detection scheme.} The generic setup configuration for the detection of SPDC photons.}
			\label{fig:DetectionScheme}
	\end{minipage}\hfill
	\begin{minipage}{0.52\textwidth}
		\centering
		\includegraphics[width=0.98\columnwidth]{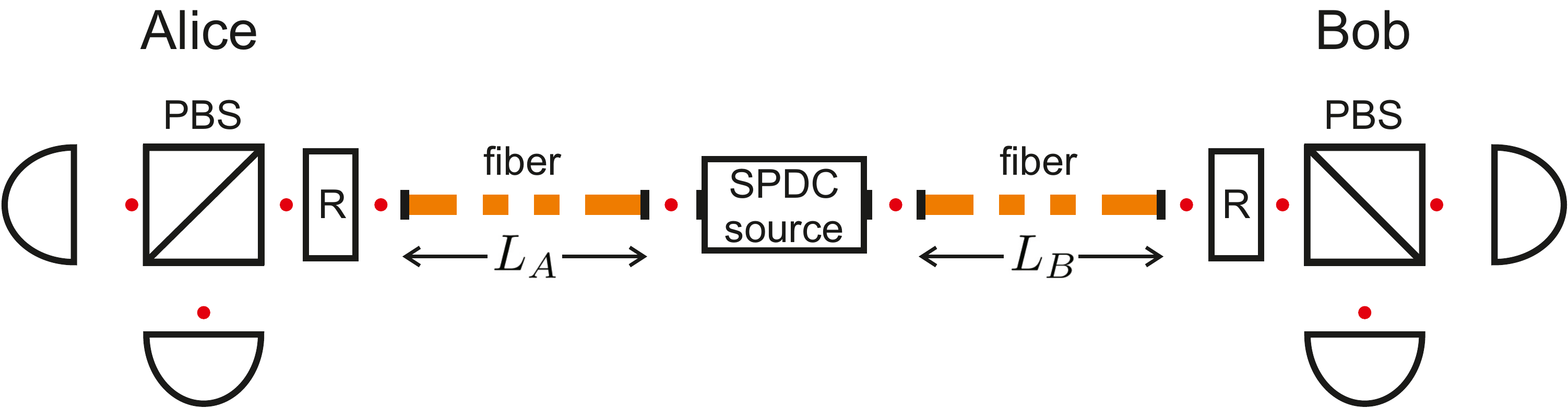}
		\caption{{\bf The QKD scheme.} A simple scheme  of discrete-variable QKD with the source of entangled photon pairs placed outside of Alice's and Bob's laboratories. R denotes polarization rotators.}
		\label{fig:QKDScheme}
	\end{minipage}
\end{figure}

In our work we consider SPDC source of photon pairs based on a nonlinear crystal parametrized by the effective phase-matching function width  $\sigma$, pumped by laser pulses of temporal width $\tau_p$. Throughout this manuscript we assume that the information on the time moments at which the laser sends pump pulses to the crystal is available to everyone interested. Photons generated by the SPDC source are subsequently transfered to single-photon detectors through quantum channels of length $L_A$ and $L_B$, characterized by the group velocity dispersion (GVD) equal to $2\beta_A$ and $2\beta_B$, respectively. This scheme is illustrated in \figref{fig:DetectionScheme}. We use the following notation to simplify the analytical calculations: $D_X\equiv\beta_XL_X$ (for $X=A,B$). To show the potential to improve the performance of QC schemes by optimizing the SPDC source we also consider a basic setup configuration for the realization of BB84 protocol in the entanglement-based variant, presented in \figref{fig:QKDScheme}. All the subsequent figures shown in this manuscript are made for $\beta_A=\beta_B=-1.15\times10^{-26}\,\mathrm{s}^2/\mathrm{m}$. For the standard wavelength of $1550\,\mathrm{nm}$ it corresponds \cite{RPPhotEnc} to the dispersion value of $18\,\mathrm{ps}/(\mathrm{nm}\cdot\mathrm{km})$, that is typical for single-mode fibers (SMFs) \cite{Corning}, which are the most common telecommunication channels utilized in practical QC schemes. During the QKD security analysis we also assume the typical value of their attenuation coefficient: $\alpha_A=\alpha_B=0.2\,\mathrm{dB}/\mathrm{km}$.

\section*{Results}

\begin{figure}[tbp]
	\centering
	\includegraphics[width=0.95\columnwidth]{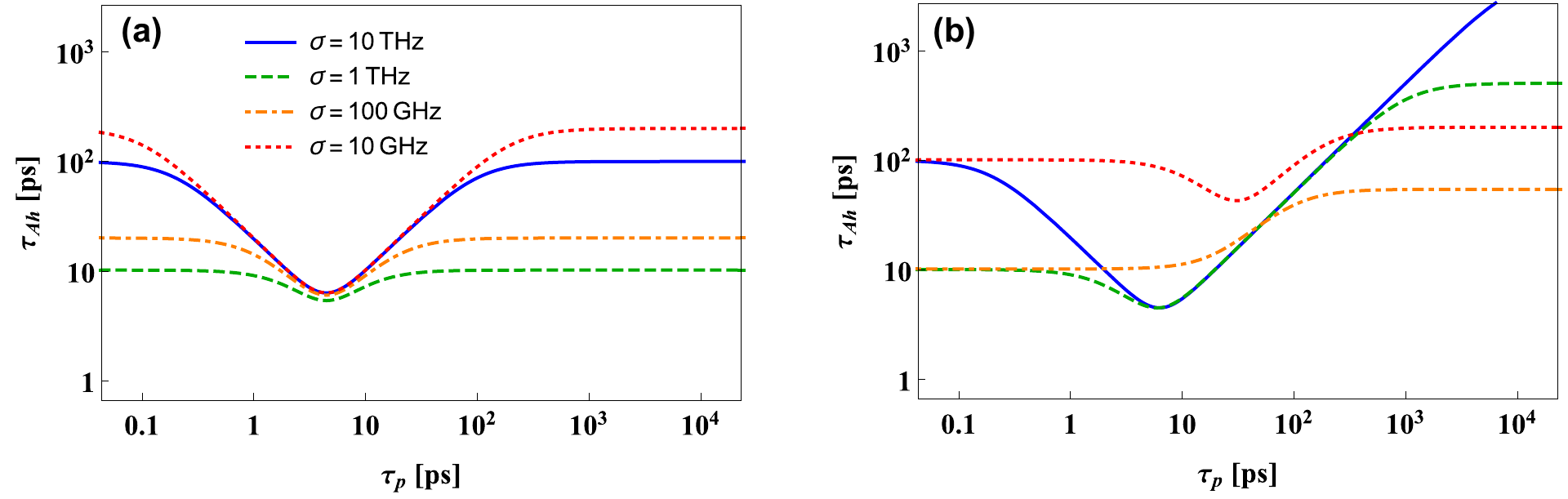}
	\caption{{\bf The dependence of $\tau_{Ah}$ on the pump laser.} The temporal width of the heralded photon A, $\tau_{Ah}$, plotted as a function of the duration time of a pump laser pulse for the case when two standard single-mode fibers (SMFs) of length $L_A=1\,\mathrm{km}$ and (a) $L_B=1\,\mathrm{km}$, (b) $L_B=100\,\mathrm{km}$ are placed between the crystal and single-photon detectors. The legend corresponds to both panels.}
	\label{fig:OptimalPump}
\end{figure}

\subsection*{Optimization of temporal widths}
As has already been stated in the Introduction, in the context of QC applications it is desirable for the temporal widths of SPDC photons to be as narrow as possible. Therefore, a natural question is: what are the optimal values of the source parameters, $\tau_p$ and $\sigma$, for which the temporal widths of SPDC photons, written explicitly in the Methods section (formulas (\ref{eq:tauA}) and (\ref{eq:tauAh})), are the lowest? In practice it is much easier to calibrate the temporal width of pump laser pulses than to modify the effective phase-matching function for the nonlinear crystal, since the latter usually requires replacing the crystal itself. Therefore, we first consider the situation in which the experimenter can only change the  pump laser utilized by the SPDC source, while the crystal is fixed. 

In this case the temporal width of the unheralded photon A, $\tau_A$, reaches its lowest value, equal to $\tau_A^\mathrm{low}=(2+|D_A|\sigma^2)/(2\sigma)$, for $\tau_p=\sqrt{2|D_A|}$. Since $\tau_A$ dos not depend on $D_B$, the above result is identical for the symmetric and asymmetric setup configurations. In the symmetric case also the temporal width of the heralded photon A, $\tau_{Ah}$, reaches its minimum for the same value of $\tau_p$. It reads: $\tau_{Ah}^\mathrm{low, sym}=[2|D_A|(D_A^2\sigma^4+4)/(|D_A|\sigma^2+2)^2]^{1/2}$. On the othe hand, the optimization of $\tau_{Ah}$ over $\tau_p$ for the asymmetric setup configuration is much more complicated. In this case the function $\tau_{Ah} (\tau_p)$ does not always have a global minimum and the conditions for its existence heavily depend on the relationship between $D_A$, $D_B$ and $\sigma$. The details of this dependence can be found in the Methods section, along with the derivation of the above formulas for $\tau_A^\mathrm{low}$ and $\tau_{Ah}^\mathrm{low, sym}$.

The examples of the relationship $\tau_{Ah} (\tau_p)$ for the symmetric and highly asymmetric setup configurations can be seen in \figref{fig:OptimalPump}. In  the symmetric case, presented in panel (a), the function $\tau_{Ah} (\tau_p)$ has a well-defined minimum for any $\sigma$. Its value depends on the effective phase-matching function width relatively weekly, while the value of $\tau_p$ for which this minimum is reached is independent of $\sigma$. The situation is much different in the highly asymmetric case, illustrated in panel (b). Here both the optimal value of $\tau_p$ and the minimal value of $\tau_{Ah}$ significantly depend on the effective phase-matching function width. Moreover, for $\sigma=100\,\mathrm{GHz}$ (corresponding to $0.13\,\mathrm{nm}$ at  $1550\,\mathrm{nm}$ in terms of wavelength) none of the conditions for the existence of the global minimum of $\tau_{Ah}$ is fulfilled. In this situation $\tau_{Ah} (\tau_p)$ is monotonically increasing function (see the orange dot-dashed line). 
It can also be seen that in the asymmetric scenario the comparison between the functions of $\tau_{Ah} (\tau_p)$ plotted for different $\sigma$ heavily depends on $\tau_p$. For example, while for very short pump pulses the value of $\tau_{Ah}$ calculated for $\sigma=1\,\mathrm{THz}$ is much smaller than for $\sigma=10\,\mathrm{GHz}$, it is the opposite for large $\tau_p$. The situation like this cannot be seen in the symmetric case.

\begin{figure}[tbp]
\centering
\includegraphics[width=0.95\columnwidth]{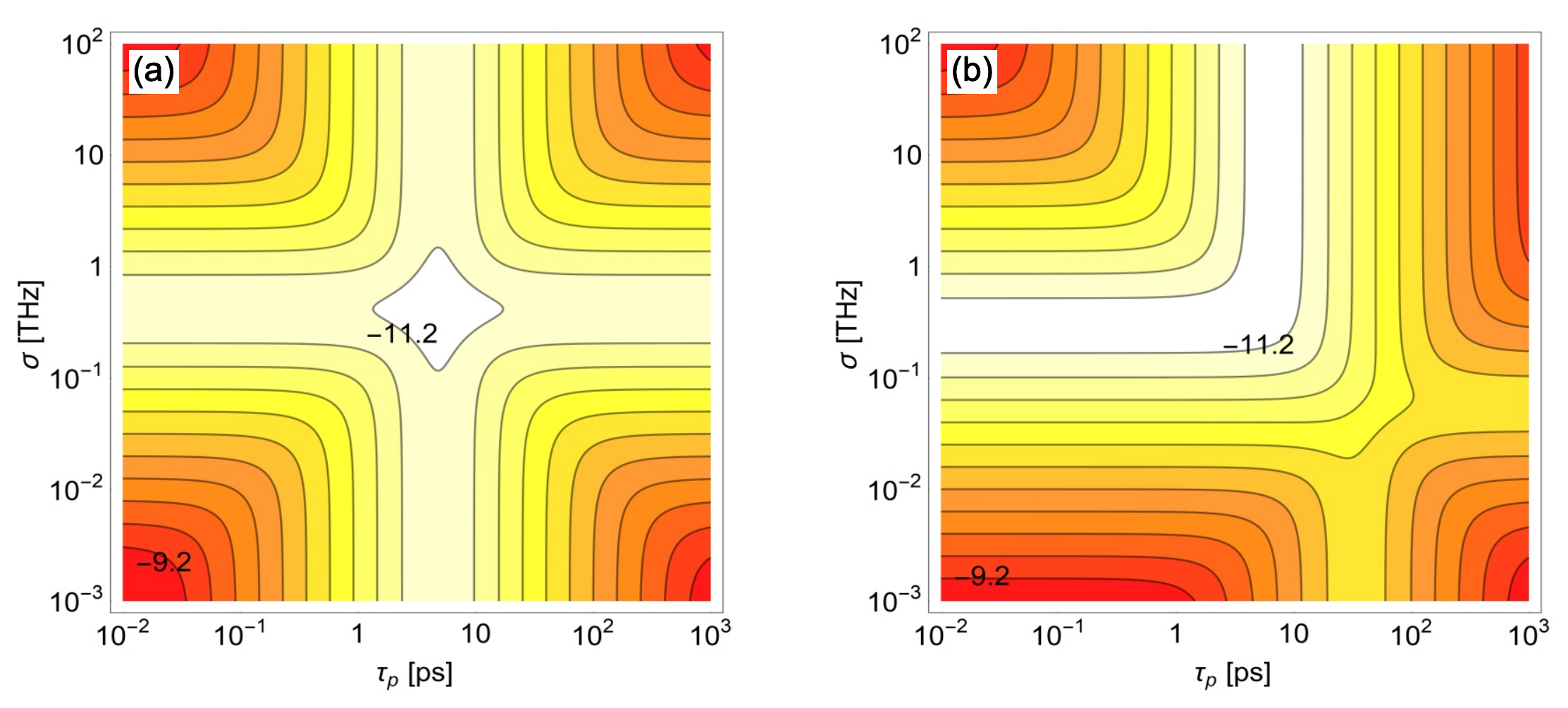}
\caption{{\bf The dependence of $\tau_{Ah}$ on the pump laser and the nonlinear crystal.} Logarithm of the temporal width of the heralded photon A, $\tau_{Ah}$, at the entrance to the detector, plotted as a function of the duration time of the pump laser pulse, $\tau_p$, and the effective phase-matching function width of the nonlinear crystal, $\sigma$, for the case when the source is connected with the detectors by SMFs of length $L_A=1\,\mathrm{km}$ and (a) $L_B=1\,\mathrm{km}$ or (b) $L_B=100\,\mathrm{km}$. The contours shown in both plots represent values from $\log_{10}\tau_{Ah}=-11.2$ to $\log_{10}\tau_{Ah}=-9.2$ with constant $0.2$ spacing.}
\label{fig:GeneralOpt}
\end{figure}

Contrary to the scenario when the nonlinear crystal is fixed, full optimization of a SPDC source over the parameters $\tau_p$ and $\sigma$ cannot be done analytically in the general case. Nevertheless, it can be performed for the symmetric setup configuration, when $D_A=D_B\equiv D$. This task has already been done in our previous paper \cite{Sedziak2017}, where it was shown that in the symmetric case the optimal values of the SPDC source parameters are $\tau_p^\mathrm{sym}=\sqrt{2|D|}$ and $\sigma^\mathrm{sym}=\sqrt{2/|D|}$. For these numbers the function $\tau_{Ah}$ reaches its absolute minimum, equal to $\tau_{Ah}^\mathrm{sym}=\sqrt{2|D|}$. In the symmetric case $\tau_{Ah}$ exhibits high symmetry both as a function of $\tau_p$ and $\sigma$. It can be seen in \figref{fig:GeneralOpt}(a), where the temporal width of the photon A is plotted for $L_A=L_B=1\,\mathrm{km}$. For comparison, in  \figref{fig:GeneralOpt}(b)  we plot $\tau_{Ah}$ for the highly asymmetric case of $L_A=1\,\mathrm{km}$ and $L_B=100\,\mathrm{km}$.

\subsection*{Dependence on the length of the heralding arm}
For the asymmetric QC scheme it is possible to reduce the temporal width of SPDC photons propagated through one of its arms by introducing a proper amount of dispersion to the other arm (\emph{e.g.} by adjusting its length). This can have positive effect on the performance of QC protocols in some setup configurations, as has already been shown in the context of the asymmetric QKD scenario \cite{Lasota2018}. However, the framework used in the aforementioned work was based on the analysis of spectral correlation and generated photon widths. It gave general insight into the physical mechanisms yielding the optimal performance of the QKD scheme, but must be reformulated to be directly related to the typical experimental scenario. Here we accomplish this goal using the parameters $\sigma$ and $\tau_p$.

\begin{figure}[h]
	\centering
	\includegraphics[width=0.45\columnwidth]{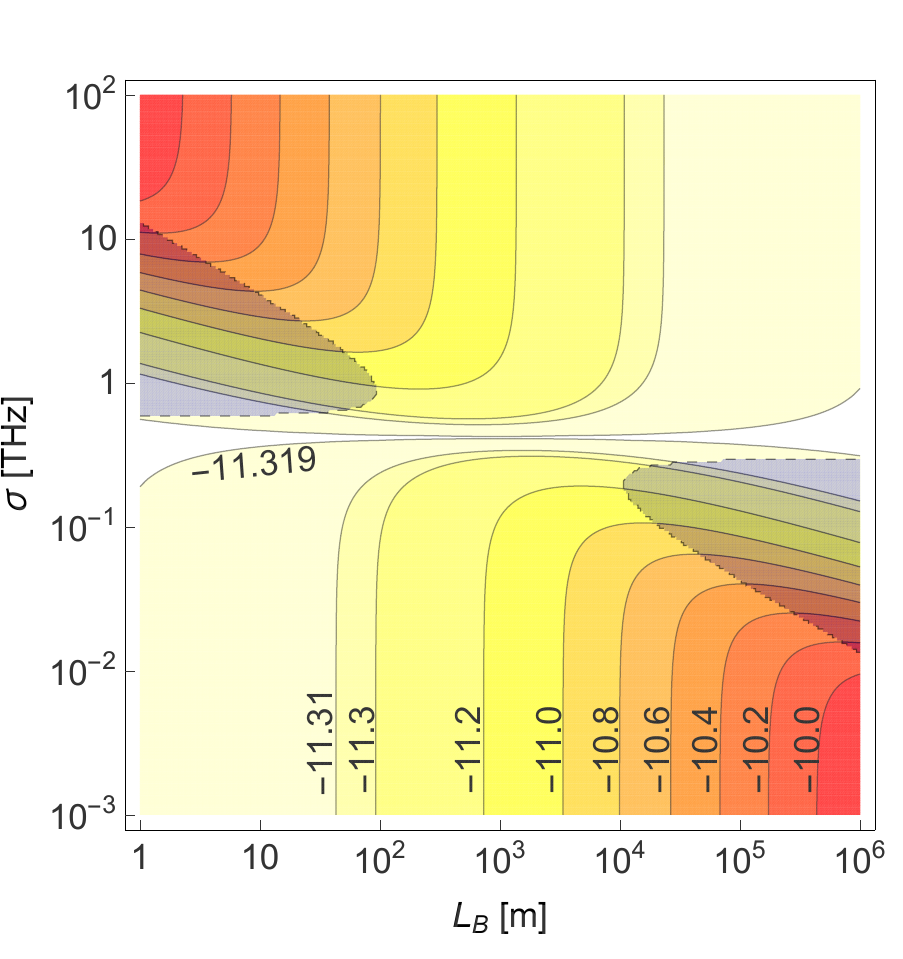}
	\caption{{\bf The dependence of $\tau_{Ah}$ on nonlinear crystal and the length of the heralding arm.} Logarithm of the temporal width of the heralded photon A at the entrance to the detector, $\tau_{Ah}$, shown as a function of the length of the heralding SMF quantum channel, $L_B$, and the effective phase-matching function width of the nonlinear crystal, $\sigma$, plotted for the case when the source is connected with the detector A by another SMF quantum channel of length $L_A=1\,\mathrm{km}$. For every pair of values $(L_B,\sigma)$ the calculated temporal width has been optimized over the pump laser pulse duration $\tau_p$. The overshadowed area near the left [right] edge of the figure corresponds to the range of $(L_B,\sigma)$, for which the optimal value of $\tau_{Ah}$ is reached for $\tau_p\rightarrow\infty$ [$\tau_p\rightarrow 0$]. For other combinations of $L_B$ and $\sigma$ the optimal value of $\tau_p$ is given by the formula (\ref{eq:OptimalTauP}). The spacing between the neighboring contours becomes smaller than $0.2$ for $\log_{10}\tau_{Ah}<-11.2$ in order to better illustrate how this function behaves near its minimum.}
	\label{fig:OptOverDistance}
\end{figure}

In \figref{fig:OptOverDistance} it can be seen how the temporal width $\tau_{Ah}$, optimized over the duration time of the pump laser pulses, changes with different values of $L_B$ and $\sigma$. This picture shows that the conditions for the function $\tau_{Ah} (\tau_p)$ to have a global minimum, derived in the Methods section, are always fulfilled when the channel parameters, $D_A$ and $D_B$, are of the same order of magnitude. Only for highly asymmetric schemes this function can be minimized asymptotically for $\tau_p\rightarrow\infty$ or $\tau_p\rightarrow0$. Furthermore, it is interesting to note that if the effective phase-matching function width is significantly larger than its optimal value, $\sigma_\mathrm{opt}$, extending the length of the heralding arm always leads to the reduction of $\tau_{Ah}$. This effect is more prominent for smaller distances, while for $L_B\rightarrow\infty$ the temporal width of the heralded photon asymptotically decreases to a fixed value. On the other hand, when $\sigma<\sigma_\mathrm{opt}$, extending the heralding arm has the opposite effect on $\tau_{Ah}$ to the one described above. Therefore, one can conclude that for the asymmetric QKD scheme the maximal security distance in one arm can be extended by introducing more dispersion to the other arm, as long as $\sigma>\sigma^\mathrm{opt}$. This conclusion is similar to the one stated in our previous work \cite{Lasota2018}. However, it is important to underline that contratry to the aforementioned paper, here the global time reference, \emph{i.e.} the timing information on the pump pulses, is assumed to be known by Alice and Bob, as has already been stated in the Introduction. Consequently, it can potentially have much broader practical application than it was previously thought\cite{Lasota2018}. 

Since the simplest way of introducing more dispersion to the heralding arm is to use longer telecommunication fiber, one should be aware of the fact that such action would always reduce the probability of registering the heralding photon. Therefore, one should avoid it as long as the length of the heralded arm is short enough to provide the QKD security even without introducing additional dispersion, as it would unnecessarily descrease the key generation rate. Only when the length of the heralded arm is indeed too long for the security of the traditional setup configuration, lenghtening the heralding fiber may provide positive results for the participants of the QKD protocol.

\begin{figure}[t]
	\centering
	\begin{minipage}{0.48\textwidth}
		\centering
		\includegraphics[width=\textwidth]{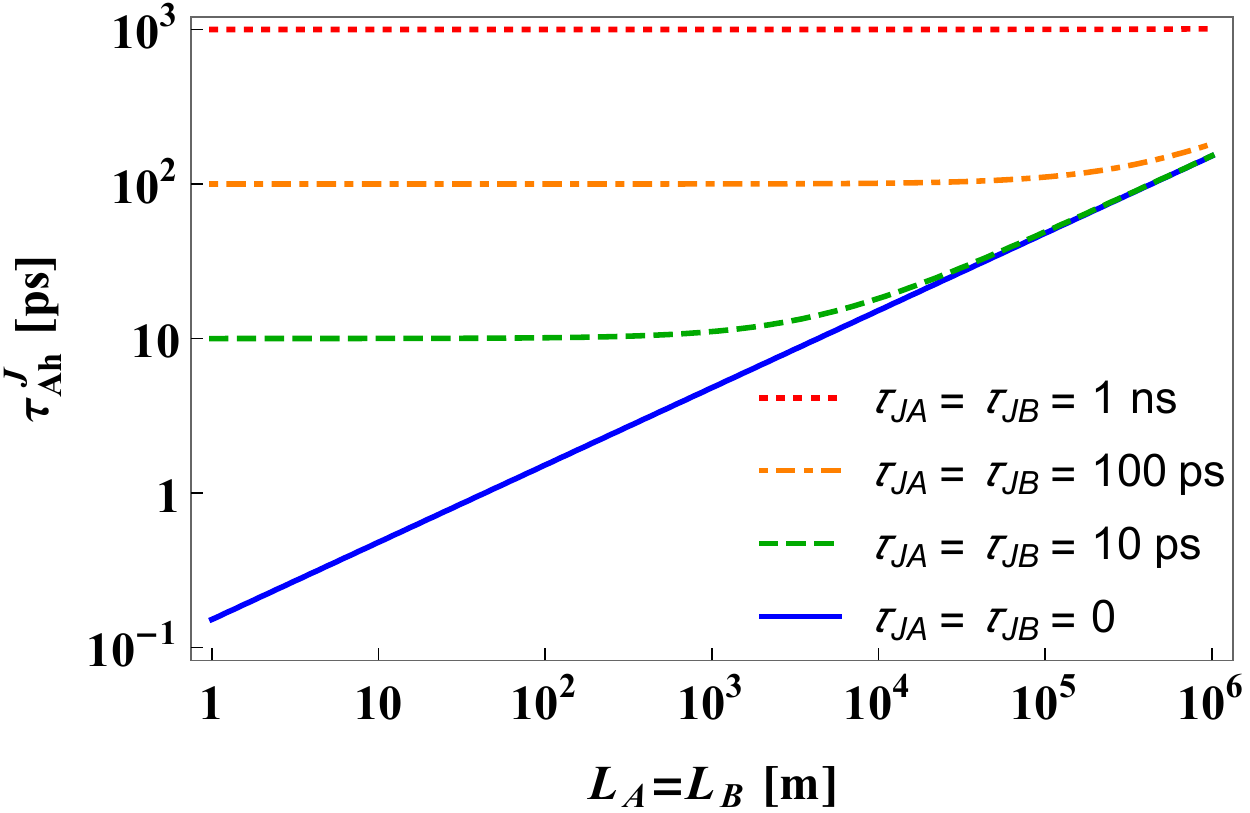}
		\caption{{\bf The influence of non-zero timing jitter on $\tau_{Ah}$.} Temporal width of the heralded photon, $\tau_{Ah}^J$, given by the formula (\ref{eq:jitter}),optimized over the SPDC source parameters $\sigma$ and $\tau_p$, plotted as a function of the length of SMFs separating the source and the photon detection systems in the case of symmetric setup configuration. The detectors' timing jitter is defined as the standard deviation of the detection time probability function.} \label{fig:JitterInfluence}
	\end{minipage}\hfill
	\begin{minipage}{0.48\textwidth}
		\centering
		\includegraphics[width=\textwidth]{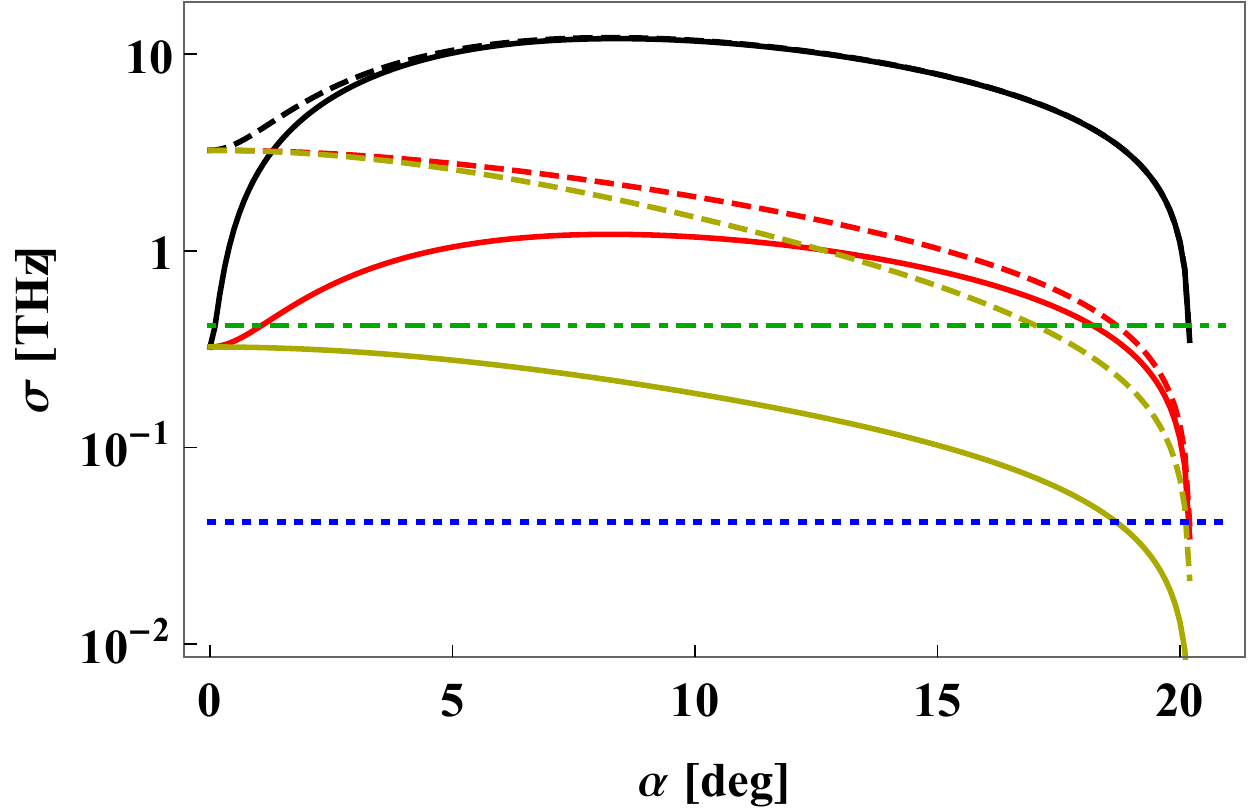}
		\caption{{\bf Realistic values of $\sigma$ for BBO crystal.} The effective phase-matching function width, $\sigma$, calculated for $775\,\mathrm{nm}\rightarrow1550\,\mathrm{nm}+1550\,\mathrm{nm}$ type I SPDC process, plotted as a function of the angle $\alpha$ between the central propagation directions for the pump photons and the generated photons. The plots are made for the crystal length  equal to $L_\mathrm{cryst}=1\,\mathrm{cm}$ (solid lines) and $L_\mathrm{cryst}=1\,\mathrm{mm}$ (dashed lines), and for the following widths of the transverse spatial modes collected by the SMFs, $W_f$: $10\,\mathrm{\mu m}$ (black lines), $100\,\mathrm{\mu m}$ (red lines), $1\,\mathrm{mm}$ (yellow lines). Blue dotted (green dot-dashed) line correspond to the optimal value of $\sigma$, calculated for the symmetric QC setup with SMF quantum channels of $100\,\mathrm{km}$ ($1\,\mathrm{km}$) length.}
		\label{fig:EPMFforBBO}
	\end{minipage}
\end{figure}

\subsection*{Dependence on the detector timing jitter} 
It can be seen in \figref{fig:OptimalPump}--\figref{fig:OptOverDistance} that if the SMF connecting the source with the detector A is of the order of $1\,\mathrm{km}$, the temporal width $\tau_{Ah}$ can be reduced even below the level of $10\,\mathrm{ps}$. This value is comparable with the timing jitter of the best currently existing single-photon detectors \cite{Shcheslavskiy2016,Yan2012,Tosi2009,Divochiy2018}. In order to estimate the range of fiber lengths for which non-zero jitter can have significant influence on the temporal widths of SPDC photons, in \figref{fig:JitterInfluence} we compare the temporal widths of the heralded photon A, $\tau_{Ah}^J$, optimized over the source parameters, $\sigma$ and $\tau_p$, calculated as a function of the propagation distance for a few different values of the timing jitter $\tau_{JA}$ and $\tau_{JB}$. The mathematical formula for $\tau_{Ah}^J$ is derived in the Methods section. The plots in \figref{fig:JitterInfluence} are made for the symmetric QC scheme. As one can see there, if $L_A$ and $L_B$ are shorter than a few kilometers, the jitter significantly influences $\tau_{Ah}^J$ even if it is much smaller than in the case of the state-of-the-art single-photon detectors. Therefore, to exploit the full potential of the optimization method presented in this paper further development of photon detection technology will be needed. At present, however, it is certainly possible to make the influence of detection jitter negligible if the propagation distance is of the order of tens of kilometers or more. To conclude, the results of our investigation, presented in \figref{fig:JitterInfluence}, indicate that if the experimenter wants to fully optimize the short-distance QC scheme, the detection jitter of realistic single-photon detectors can become one of the most important factors. On the other hand, for long-distance communication schemes the jitter can be safely neglected.

\subsection*{Realistic values of the effective phase-matching function width}
In principle, in the case of any specific QC setup configuration, using the optimization rules presented in this paper allows the experimenter to easily find the most favourable values of a pump laser pulse duration and an effective phase-matching function width. However, one may wonder if these theoretically optimal values would be achievable for realistic SPDC sources. It is much easier to answer this question in the context of the pump laser pulse duration, owing to the variety of commercially available lasers, ranging from the CW to femtosecond ones. Since the optimal value of $\tau_p$ generally grows with the propagation distance and already for $L_A=L_B=1$ m it is approximately equal to $150\,\mathrm{fs}$, one can safely say that the theoretically optimal pump laser pulse duration should be achievable for basically every realistic QC scheme.

Performing similar analysis in the context of the effective phase-matching function width associated with different kinds of nonlinear crystals is much more complexed. The value of $\sigma$ depends not only on the type of nonlinear material, but also on several other parameters such as the crystal length or its optical axis orientation \cite{Gajewski2016}. However, in order to get some intuition in this matter, we analyzed here a specific case of BBO crystal cut for degenerate type I SPDC process, in which $775\,\mathrm{nm}$ pump photons are converted to pairs of $1550\,\mathrm{nm}$ photons. The results of our investigation are presented in \figref{fig:EPMFforBBO}, where the effective phase-matching function width was plotted as a function of the angle $\alpha$ between the central propagation directions of the pump photons and the generated photons. The calculations were made for several different values of the crystal length, $L_\mathrm{cryst}$, and the width of transverse spatial mode collected by the SMFs, $W_f$. Additionally, the optimal values of $\sigma$ for symmetric QC setup configuration using SMFs of length $L=1\,\mathrm{km}$ and $L=100\,\mathrm{km}$ were indicated in this figure for comparison.

The most important conclusion that can be drawn from \figref{fig:EPMFforBBO} is that for the source based on BBO crystal analyzed here the theoretically optimal values of the effective phase-matching function width can be very difficult to obtain in most practical situations. This goal seems to be especially hard to achieve for the case of collinear source configuration, \emph{i.e.} when $\alpha=0$, which is often the most convenient one in practice. In this situation, even when using exceptionally long BBO crystals, one may hope to obtain $\sigma^\mathrm{opt}$ width only for short-distance QC schemes. In principle, smaller widths of the effective phase-matching function can be get when the values of $W_f$ are sufficiently large and the BBO crystal is cut to emit pairs of photons at broad angle from the direction of propagation of the pump laser pulses. However, this kind of SPDC source would be significantly more difficult to construct. Moreover, its pair production efficiency and heralding efficiency would most likely be much smaller than for the case of collinear configuration. This would negatively affect the performance of many QC protocols \cite{Kolenderski2009}.  

\subsection*{An example of application: quantum key distribution}

\begin{figure}[t]
\centering
\includegraphics[width=0.95\columnwidth]{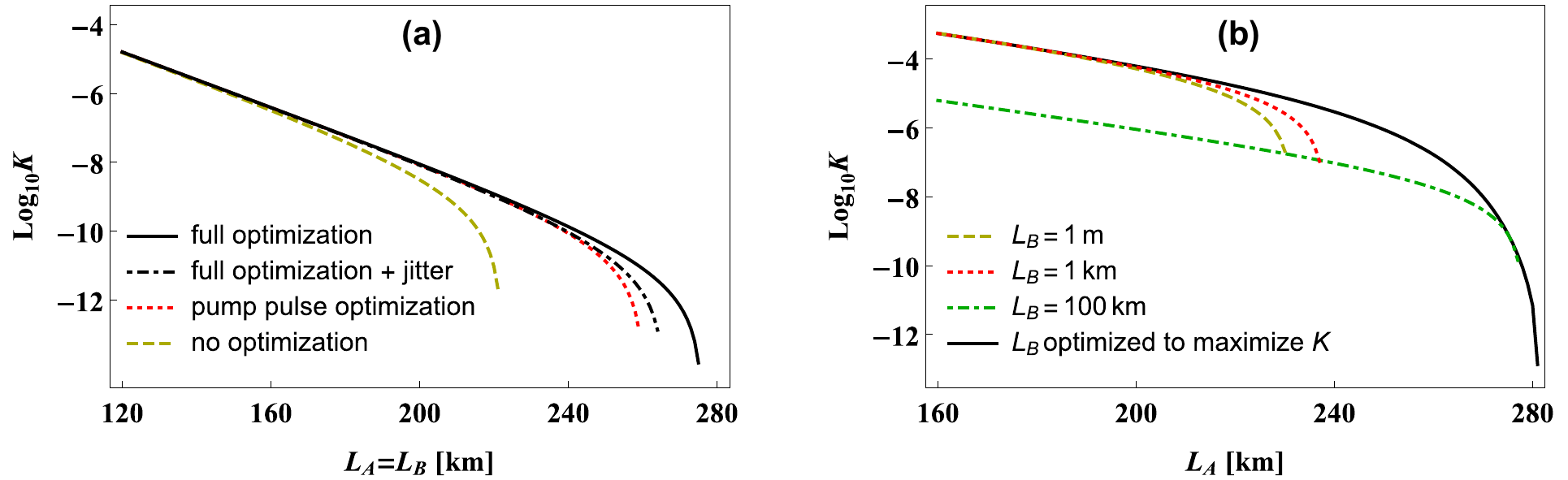}
\caption{{\bf Key generation rate.} The lower bound for the key generation rate, $K$, plotted as a function of the length $L_A$ of the SMF used to connect the SPDC source with the laboratory of Alice, $L_A$, for the BB84 protocol performed in (a) symmetric, (b) asymmetric version of the QKD scheme presented in \figref{fig:QKDScheme}. In panel (a) the plots are calculated for the following values of the source parameters: $\sigma=1\,\mathrm{THz}$ and $\tau_p=1\,\mathrm{ns}$ (dashed yellow line), $\sigma=1\,\mathrm{THz}$ and $\tau_p=\sqrt{2|\beta|L}$ (dotted red line), $\sigma=\sqrt{2/(|\beta|L)}$ and $\tau_p=\sqrt{2|\beta|L}$ (solid black line). The three aforementioned curves are drawn assuming ideal single-photon detectors with no timing jitter. Additionally, dot-dashed black line illustrates the lower bound for the key generation rate calculated in the case when the jitter of all the detectors utilized by Alice and Bob is $100\,\mathrm{ps}$, while the source parameters are $\sigma=\sqrt{2/(|\beta|L)}$ and $\tau_p=\sqrt{2|\beta|L}$. In panel (b) all of the plots are made for $\sigma=1\,\mathrm{THz}$, while the values of $\tau_p$ are numerically optimized and the jitter is assumed to be zero.}
\label{fig:QKDSecurityBoth}
\end{figure}

The potential of the presented method for the optimization of a SPDC source for its use in QC applications can be seen in \figref{fig:QKDSecurityBoth}(a), where we plot the lower bound for the key generation rate that can be obtained from the realization of BB84 protocol in the symmetric version of the setup configuration schematically illustrated in \figref{fig:QKDScheme}. We analyzed the cases of (i) non-optimized source with $\sigma=1\,\mathrm{THz}$ and $\tau_p=1\,\mathrm{ns}$, (ii) the source with the same $\sigma$, but optimized over the value of $\tau_p$ and (iii) the fully optimized SPDC source. The value of the pump laser pulse duration used in the case (i) is roughly the same as in one of the rare experimental realizations of long-distance QKD with SPDC sources \cite{Sun2014}. Since the authors of the aforementioned paper did not calculate the effective phase-matching function width for their crystal, we decided to use a typical value here. Technical details of the security analysis can be found in the Methods section.

It can be seen in \figref{fig:QKDSecurityBoth}(a) that in principle by fully optimizing the source the maximal security distance for the analyzed scheme can be extended by almost sixty kilometers for each of the two existing quantum channels, which is around 30\% compared to the non-optimized case. Moreover, even partial optimization of the source, just over the pump laser pulse duration, can provide the legitimate participants of the BB84 protocol with about $20\%$ of additional security distance. It is also important to notice, that the results plotted in \figref{fig:QKDSecurityBoth}(a) do not change considerably if we assume that Alice and Bob use single-photon detectors characterized by detection timing jitter of $\tau_{JA}=\tau_{JB}=100\,\mathrm{ps}$, which is well above the best achievable value for the modern devices \cite{Tosi2009,Yan2012,Shcheslavskiy2016}. In this situation the maximal security distance is shortened only by a few kilometers compared to the case with ideal single-photon detectors. This result is consistent with the jitter influence analysis presented earlier in this work.

The results shown in \figref{fig:QKDSecurityBoth}(a) were obtained for the security analysis of the symmetric version of the QKD setup. In \figref{fig:QKDSecurityBoth}(b) we present the results concerning more general situation, in which the two SMFs connecting the SPDC source with Alice and Bob are not of the same length. In this part of our work we specifically focus on checking how changing the length of Bob's fiber can influence the maximal security distance between the source and Alice. This investigation is motivated by the possibility of decreasing the temporal width of the heralded SPDC photon by extending the distance between the source and the heralding detector, discussed before. While such possibility is available only when $\sigma>\sigma^\mathrm{opt}$ (see \figref{fig:OptOverDistance} (a)), the plots shown in \figref{fig:EPMFforBBO} strongly suggest that this requirement can be fulfilled in most practical situations. As can be seen in \figref{fig:QKDSecurityBoth}(b), the maximal security distance between the source and Alice can be increased by several tens of kilometers just by optimizing the value of $L_B$ for any given $L_A$, instead of fixing it on some short level.

\section*{Discussion}

In this work we performed theoretical optimization of SPDC photon pairs for QC schemes with two dispersive quantum channels of arbitrary lengths. It was done over the pump laser pulse duration and the effective phase-matching function width of nonlinear crystal. We derived an analytical formula for the best setting of the pump laser for a given crystal in the most general case.
Moreover, we performed full numerical optimization of a SPDC source, demonstrating the possibility to further refine the performance of quantum protocols. 
We also showed that the temporal width of a SPDC photon can be minimized in one of two possible ways, depending on the exact value of the effective phase-matching function width: either by increasing the dispersion in the quantum channel or by decreasing it. The first (second) of these possibilities is available when the effective phase-matching function width is larger (smaller) than its optimal value.  

To compare theoretical predictions of our work with capabilities of realistic SPDC sources we investigated the source based on BBO crystal, designed for type I SPDC process generating pairs of $1550\,\mathrm{nm}$ photons. For such source we performed analytical estimation of the effective phase-matching function width. It should be noted here that precise calculation of this parameter can be done only numerically and is beyond the scope of this analysis. The obtained results suggests that for most QC schemes the achievable value of the effective phase-matching function width would be significantly larger than the theoretically optimal one. While in some situations the optimal value could be achieved, it would often require relatively large angles between the pump laser pulse direction and the propagation directions of the generated photons. However, such setup configuration would negatively affect the efficiency of SPDC source. The above consideration raises the question in what situations it would be more beneficial to abandon the full optimization of the SPDC source based on BBO crystal and use the collinear configuration to produce pairs of photons, and when it would be better to push for the full optimization at the expense of efficiency of the source. Further analysis of this problem would be required to reliably answer such question. Moreover, similar investigation performed for other types of nonlinear crystals would be necessary.

To demonstrate the potential for improving the performance of QC protocols by optimizing SPDC source, we analyzed simple entanglement-based QKD scheme. We showed that the maximal secure communication distance can be significantly increased just by properly adjusting the pump laser. Furthermore, if the full optimization of the source is possible, the improvement may even reach $30\%$ in comparison with the practical non-optimized scheme. We also showed that in realistic cases the detection timing jitter reduces the maximal security distance by no more than a few kilometers. 

The method for improving the QKD security presented in this paper is based on the reduction of the amount of noise registered by single-photon detectors during the protocol. Therefore, its effectiveness could be significantly smaller if the decrease of the key generation rate to zero at the maximal security distance was mainly caused by some other factors than the reduction of signal-to-noise ratio below the critical level. Specifically, if the SPDC source is used in a prepare-and-measure type of QKD setup configuration, the security of such scheme could be higly dependent on the probability for producing more than one pair of photons, which is always non-zero in realistic situations. However, the damaging influence of the multipair generation events on QKD security can be efficiently reduced by using decoy-pulse method \cite{Hwang2003}, which greatly limits the possibility to attack multiphoton pulses by a potential eavesdropper. While most of the recent record-breaking long-distance realizations of QKD protocols reported in the literature have been implemented using weak coherent pulses and decoy-pulse method \cite{Korzh2015,Yin2016,Boaron2018}, many papers suggest that heralded single-photon sources could potentially be better for this task \cite{Wang2007,Wang2013a,Zhu2017}. This notion can be supported by taking into account the recent advances in the field of heralding efficiency of the SPDC sources \cite{Pomarico2012,Ramelow2013,Kaneda2016}. While the strong temporal broadening of the generated signals has always been a serious obstacle for using these sources in fiber-based long-distance communication, their optimization method presented here allows to overcome this important problem.

Taking into account the above considerations, the noise registered by the measurement systems during the key generation procedure appears to be much bigger issue for long-distance QC than the aforementioned imperfection of photon pair sources. Since in our work we considered dark counts as the only source of noise, one can expect that the SPDC source optimization method can provide even better results in more realistic cases. It seems to be especially promising for the QKD performed in commercial fibers populated by strong classical signals, where the level of channel noise caused by those signals is typically very high \cite{Eraerds2010}. Our results can be particularly useful in the case of asymmetric QKD scheme in which the distance between one of the parties and the source is relatively small and the goal is to maximally extend the security length of the quantum channel connecting the source with the other party. A good example of such scenario can be found when considering a communication between a single individual user and a distant node in a multilevel quntum network with several access networks connected to the central backbone \cite{Elkouss2013,Ciurana2014}. 
Then, the maximal security distance between two separate access networks could be substantially increased by introducing more dispersion to the quantum channels connecting the individual users with their respective central nodes, as we also demonstrated in this work.

\section*{Methods}

\subsection*{Temporal widths of SPDC photons}
The spectral wavefunction of the pairs of photons produced by an SPDC source can be written in the following approximate form \cite{Lutz2014,Gajewski2016}:
\begin{equation}
\phi(\nu_A,\nu_B)\propto\exp\left(-\frac{\left(\nu_A-\nu_B\right)^2}{\sigma^2}-\frac{\left(\nu_A+\nu_B\right)^2\tau_p^2}{4}\right),
\label{eq:SpectralWavefunction}
\end{equation}
where $\nu_A$, $\nu_B$ are frequency detunings from the respective central frequencies. To calculate the temporal wavefunction of the pair of SPDC photons after their propagation through the dispersive media we utilize the following formula \cite{Sedziak2017}:
\begin{equation}
\psi_{D_AD_B}\left(t_A,t_B\right)=\frac{1}{4\pi i\sqrt{D_AD_B}}\int\mathrm{d}t_A'\int\mathrm{d}t_B'\exp\left(\frac{i\left(t_A-t_A'\right)^2}{4D_A}+\frac{i\left(t_B-t_B'\right)^2}{4D_B}\right)\psi\left(t_A',t_B'\right),
\label{eq:psi}
\end{equation}
where $\psi\left(t_A',t_B'\right)$ denotes the initial temporal wavefunction. It can be obtained from $\phi(\nu_A,\nu_B)$ through Fourier transform. 

Without any loss of generality we focus on calculating the temporal width of the photon entering the detector A (photon A) in \figref{fig:DetectionScheme}. 
If an experimenter knows nothing about the detection time of the other photon (photon B), the probability distribution function for the detection time of photon A can be calculated as the marginal distribution $p_A(t_A)=\int\mathrm{d}t_B|\psi_{D_AD_B}(t_A,t_B)|^2$. In this case the temporal width of photon A reads:
\begin{equation}
\tau_A=\frac{\sqrt{\left(\tau_p^2+D_A^2\sigma^2\right)\left(4+\sigma^2\tau_p^2\right)}}{2\sigma\tau_p}
\label{eq:tauA}
\end{equation}
On the other hand, if the detection time of photon B is known to be $T_B$, the probability distribution function for the detection time of photon A takes the form of $p_{Ah}(t_A,t_B=T_B)=|\psi_{D_AD_B}(t_A,t_B=T_B)|^2/[\int\mathrm{d}t_A|\psi_{D_AD_B}(t_A,t_B=T_B)|^2]$. Its temporal width is then given by:
\begin{equation}
\tau_{Ah}=\sqrt{\frac{16\left(\tau_p^2-D_AD_B\sigma^2\right)^2+\left(D_A+D_B\right)^2\left(\sigma^2\tau_p^2+4\right)^2}{4\left(\tau_p^2+D_B^2\sigma^2\right)\left(\sigma^2\tau_p^2+4\right)}}.
\label{eq:tauAh}
\end{equation}
The temporal width of photon B in the non-heralded and heralded case can be obtained immediately from the expressions (\ref{eq:tauA}) and (\ref{eq:tauAh}), respectively, by switching $D_A$ to $D_B$ and vice versa.

\subsection*{Optimization of the pump laser in the asymmetric case} The conditions for the function $\tau_{Ah}(\tau_p)$ to have a well-defined global minimum are very complicated in the general case. However, they can be considerably simplified if we assume that $D_A$ and $D_B$ have the same sign, which is certainly a justified assumption in realistic situations. To write them explicitly we first introduce the following notation:
\begin{equation}
\xi_{i,j}\!=\!\!\left[i\frac{D_A\!-\!D_B\!+\!j\sqrt{(D_A\!-\!D_B)^2\!-\!8D_A(D_A\!+\!D_B)}}{2D_AD_B}\right]^{1/2},
\end{equation}
\begin{equation}
\zeta_{i,j}\!=\!\!\left[i\frac{D_A\!-\!D_B\!+\!j\sqrt{(D_A\!-\!D_B)^2\!-\!8D_B(D_A\!+\!D_B)}}{D_B(D_A\!+\!D_B)}\right]^{1/2}.
\end{equation}
 Here we focus on the typical QC scheme with SMFs, in which case it is always $D_A,D_B<0$. Then the right-hand side of the expression (\ref{eq:tauAh}) reaches its minimum for
\begin{equation}
\tau_{p}^{(-)}\!=2\sqrt{-\frac{2(D_A\!+\!D_B)\!-\!\sigma^2D_B(D_A\!-\!D_B)\!+\!\sigma^4D_AD_B^2}{8\!+\!2\sigma^2(D_A\!-\!D_B)\!+\!\sigma^4D_B(D_A\!+\!D_B)}}
\label{eq:OptimalTauP}
\end{equation} 
in the three following cases: (1) when $10.7\,D_A\approx(4\sqrt{2}+5)D_A<D_B<\left[(4\sqrt{2}-5)/7\right]D_A\approx0.094\,D_A$, (2) when $D_B\leq(4\sqrt{2}+5)D_A$ and one of the inequalities $\sigma<\xi_{+1,-1}$ or $\xi_{+1,+1}<\sigma$ is true, (3) when $\left[(4\sqrt{2}-5)/7\right]D_A\leq D_B$ and one of the inequelities $\sigma<\zeta_{-1,+1}$ or $\zeta_{-1,-1}<\sigma$ is true. If none of the above sets of conditions is fulfilled, then the function $\tau_{Ah} (\tau_p)$ does not have a global minimum. If $D_B\leq(4\sqrt{2}+5)D_A$ but $\xi_{+1,-1}\leq\sigma\leq\xi_{+1,+1}$ it is monotonically increasing, meaning that the lowest temporal width of photon A is reached for $\tau_p\rightarrow 0$. On the other hand if $\left[(4\sqrt{2}-5)/7\right]D_A\leq D_B$ but $\zeta_{-1,+1}\leq\sigma\leq\zeta_{-1,-1}$ the function $\tau_{Ah} (\tau_p)$ always decreases when $\tau_p$ grows. Therefore, in this situation the lowest temporal width of photon A is reached for $\tau_p\rightarrow\infty$.

\subsection*{Temporal widths of SPDC photons when the detection timing jitter is non-zero}
When the timing jitter, $\tau_{JA}$, is non-zero the difference between the detection time of photon A, $t_A$, and the time of its arrival at the measurement system, $t_A^0$, can be described by the probability distribution function $q(t_A,t_A^0,\tau_{JA})=M_A\exp[-(t_A-t_A^0)^2/(2\tau_{JA}^2)]$,
where $M_A$ is the normalization constant. Then, the probability distribution for the detection time of this photon in the case when the detection time of photon B is unknown can be calculated as $\pi_A(t_A)=\int\mathrm{d} t_A^0\,p_A(t_A^0)q(t_A,t_A^0,\tau_{JA})$. The marginal distribution function $p_A(x)$ has already been defined in the text between the equations (\ref{eq:psi}) and (\ref{eq:tauA}). It is straightforward to check that the standard deviation of $\pi_A(t_A)$ is equal to $\tau_A^J=(\tau_A^2+\tau_{JA}^2)^{1/2}$. The above formula gives the temporal width of the non-heralded photon A for the case of non-zero jitter. 

While the value of $\tau_A^J$ depends only on the timing jitter of the detector A, the analogous temporal width of photon A found in the heralded case, $\tau_{Ah}^J$, would be influenced also by the timing jitter of the other detector, $\tau_{JB}$. In order to calculate it, one has to take the joint probability formula for the detection of photon A at the time $t_A$ and the detection of photon B at the time $t_B$, which can be derived from (\ref{eq:psi}) as $p_{AB}(t_A,t_B)=|\psi_{D_AD_B}\left(t_A,t_B\right)|^2$, and modify it to the following form: $\pi_{AB}(t_A,t_B)=\int\mathrm{d} t_A^0\int\mathrm{d} t_B^0\,p_{AB}(t_A^0,t_B^0)q(t_A,t_A^0,\tau_{JA})q(t_B,t_B^0,\tau_{JB})$. In the above formula $t_B^0$ is the arrival time of photon B to the heralding detector. The probability distribution of the detection time of photon A, conditioned on the detection of photon B at the time $T_B$, is then given by $\pi_{Ah}(t_A,t_B=T_B)=\pi_{AB}(t_A,t_B=T_B)/\int\mathrm{d} t_A\,\pi_{AB}(t_A,t_B=T_B)$. The standard deviation of the resulting function is 
\begin{equation}
\tau_{Ah}^J=\sqrt{\tau_{Ah}^2+\tau_{JA}^2+X\tau_{JB}^2},
\label{eq:jitter}
\end{equation}
where $\tau_{Ah}$ is the temporal width of the photon A calculated for zero jitter case and $X=(\tau_p^2-D_AD_B\sigma^2)^2(\sigma^2\tau_p^2-4)^2/[(\tau_p^2+D_B^{\,2}\sigma^2)^2(\sigma^2\tau_p^2+4)^2]$.

\subsection*{Effective phase-matching function width for the BBO crystal}
The approximate value of the effective phase-matching function width for a particular nonlinear crystal can be calculated by using the following formula: $\sigma=[(\delta_k^2/W_f^2+5/L^2)/\delta_\omega^2]^{1/2}$, where $W_f$ is the width of transverse spatial mode collected by the SMF and $L$ is the length of the crystal \cite{Gajewski2016}. Furthermore, by $\delta_k$ and $\delta_\omega$ we defined the partial derivatives of the phase mismatch $\Delta k_z$ over the transverse component of wave vector of the produced signal photons and their angular frequencies, respectively. In this work we are interested in type I SPDC process taking place in BBO crystal. It is a negative uniaxial crystal, which means that the pump photons have to be extraordinarily polarized, while the polarizations of signal and idler photons are always ordinary \cite{Dmitriev1999}. Assuming that the pump pulses propagate along the $z$ direction, the phase mismatch for the investigated process is given by $\Delta k_z=(\omega_s+\omega_i)n^e(\omega_s+\omega_i,\theta)/c-[\omega_s n_o(\omega_s)/c-k_{sx}^2]^{1/2}-[\omega_i n_o(\omega_i)/c-k_{ix}^2]^{1/2}$, where $\omega_s$ ($\omega_i$) is the angular frequency of the signal (idler) photon and $k_{sx}$($k_{ix}$) is its transverse wave vector component. The refractive index for pump photons depends on the angle between the Z axis and the optic axis, $\theta$, as follows \cite{Dmitriev1999}: 
\begin{equation}
n^e(\omega,\theta)=n_o(\omega)\sqrt{\frac{1+\tan^2\theta}{1+\left[n_o(\omega)/n_e(\omega)\right]^2\tan^2\theta}}.
\end{equation}
The approximate formula for the dependence of the refractive index of the ordinarily [extraordinarily] polarized photons propagating in the BBO crystal, $n_o(\omega)$ [$n_e(\omega)$], on their angular frequency can be found in Ref.\,\cite{Dmitriev1999}. The expression for $\theta$ can be obtained by solving the equation $\Delta k_z=0$.

\subsection*{QKD security analysis}
For the BB84 protocol realized in the setup configuration illustrated in \figref{fig:QKDScheme} the lower bound on the key generation rate is given by $K=p_\mathrm{exp}[1-2H(Q)]$, where $H(x)=-x\log_{2}x-(1-x)\log_2(1-x)$ is the Shannon entropy and $Q$ denotes the quantum bit error rate (QBER) in the raw key generated by the legitimate participants of the protocol \cite{Kraus2005}. In the above formula $p_\mathrm{exp}$ is the probability of accepting a given event by Alice and Bob for the process of key generation. Obviously, both $Q$ and $p_\mathrm{exp}$ depend on the duration time of the detection windows chosen by the participants of the protocol. For a single such window of width $\xi\tau_X$ the probability for registering a photon of temporal width $\tau_X$ is given by
\begin{equation}
\eta(\xi)={(2\pi)^{-1/2}}\int_{-\xi/2}^{\xi/2}dy\,\exp(-y^2/2)=\mathrm{erf}(\xi/2\sqrt{2}).
\end{equation}
On the other hand, the probability for registering a dark count in one of the two single-photon detectors can be calculated as
\begin{equation}
P_{X}(\xi)=2d\xi\tau_{X},
\label{eq:Px}
\end{equation}
where $d$ is the dark count rate for a given single-photon detector. For the calculations performed in this work we assume that $d=1\,\mathrm{kHz}$ both in the case of Alice's and Bob's detectors. Also, for every pair of the investigated values of $L_A$ and $L_B$ we separately optimize the parameters $\xi_A$ and $\xi_B$ in order to get the best possible outcomes for the legitimate parties. We consider the situation in which the dark counts are the only source of errors in the raw key. Since narrowing the detection windows reduces all the possible errors that are uncorrelated with the real signals in exactly the same way, adding such errors to the model can be easily made just by appropriate increase of $d$. On the other hand, the errors that are connected to the real signals, \emph{e.g.} polarization rotation, would only slightly change the obtained results and not in qualitative way. 

We consider the case of perfect SPDC source, always emitting a single pair of photons when the pump pulse propagates through the crystal. With this assumption the probability $p_\mathrm{exp}$ for the scheme illustrated in \figref{fig:QKDScheme} can be approximated by
\begin{eqnarray}
p_\mathrm{exp}&\approx& T_A\eta(\xi_A)T_B\eta(\xi_B)+T_A\eta(\xi_A)[1-T_B\eta(\xi_B)]P_{Bh}(\xi_B)+[1-T_A\eta(\xi_A)]T_B\eta(\xi_B)P_{Ah}(\xi_A)+\\&+&[1-T_A\eta(\xi_A)][1-T_B\eta(\xi_B)]P_A(\xi_A)P_{Bh}(\xi_B)\nonumber,
\label{eq:pexp}
\end{eqnarray}
where $T_A$ ($T_B$) is the transmittance of quantum channel connecting the SPDC source with Alice (Bob), given by $T_A=10^{-{\alpha_A L_A}/{10}}$ ($T_B=10^{-{\alpha_B L_B}/{10}}$). The probabilities for a dark count to be registered by Alice or Bob in a particular detection window, denoted by $P_{Ah}$ and $P_{Bh}$ respectively, can be calculated by inserting the expression (\ref{eq:tauAh}), and the analogous expression for the temporal width of the heralded photon B, into the formula (\ref{eq:Px}). On the other hand, in order to obtain $P_{A}$ one should use the equation (\ref{eq:tauA}) instead of (\ref{eq:tauAh}).
This probability, appearing only in the last term on the right-hand side of Eq.\,(\ref{eq:pexp}), is needed to properly account for the case when neither of the signal photons is detected by the measurement systems of Alice and Bob. In this situation one of the dark counts registered by them  has to be treated as a ``heralding'' click, while the other one is ``heralded'' by it (obviously, the exact choice does not matter here, as can be confirmed by checking that $\tau_A\tau_{Bh}\equiv\tau_B\tau_{Ah}$). On the other hand, the second and third terms on the right-hand side of Eq.(\ref{eq:pexp}) correspond to the case when only one of the two photons produced by the source causes a click in one of the measurement systems, but the event is still accepted for the key generation process due to a dark count registered in the other detection system in the appropriately narrowed time window. Finally, the first term on the right-hand side of Eq.(\ref{eq:pexp}) accounts for the desired situation in which both SPDC photons from a given pair are detected by Alice's and Bob's measurement systems.

In the case of the simplified QKD scheme considered in our work an error in the raw key can be generated only if at least one of the signal photons from a given SPDC pair is lost, but the event is still accepted for key generation. Since dark counts occur in the detectors of Alice and Bob totally randomly, in all of such situations the error probability is $50\%$. Therefore, QBER can be calculated using the expression
\begin{equation}
Q=\frac{p_\mathrm{exp}-T_A\eta(\xi_A)T_B\eta(\xi_B)}{2p_\mathrm{exp}}.
\label{eq:qber}
\end{equation}

\bibliography{BibliographySciRep}

\begin{thebibliography}{10}
\urlstyle{rm}
\expandafter\ifx\csname url\endcsname\relax
  \def\url#1{\texttt{#1}}\fi
\expandafter\ifx\csname urlprefix\endcsname\relax\def\urlprefix{URL }\fi
\expandafter\ifx\csname doiprefix\endcsname\relax\def\doiprefix{DOI: }\fi
\providecommand{\bibinfo}[2]{#2}
\providecommand{\eprint}[2][]{\url{#2}}

\bibitem{Bennett1984}
\bibinfo{author}{Bennett, C.~H.} \& \bibinfo{author}{Brassard, G.}
\newblock \bibinfo{title}{Quantum cryptography: Public key distribution and
  coin tossing}.
\newblock In \emph{\bibinfo{booktitle}{Proceedings of the IEEE International
  Conference on Computers, Systems, and Signal Processing, Bangalore, India}},
  vol.~\bibinfo{volume}{11}, \bibinfo{pages}{175--179}
  (\bibinfo{publisher}{IEEE, New York}, \bibinfo{year}{1984}).

\bibitem{Ekert1991}
\bibinfo{author}{Ekert, A.~K.}
\newblock \bibinfo{journal}{\bibinfo{title}{Quantum cryptography based on bell
  theorem}}.
\newblock {\emph{\JournalTitle{Phys. Rev. Lett.}}}
  \textbf{\bibinfo{volume}{67}}, \bibinfo{pages}{661--663}
  (\bibinfo{year}{1991}).

\bibitem{Hillery1999}
\bibinfo{author}{Hillery, M.}, \bibinfo{author}{Bu\v{z}ek, V.} \&
  \bibinfo{author}{Berthiaume, A.}
\newblock \bibinfo{journal}{\bibinfo{title}{Quantum secret sharing}}.
\newblock {\emph{\JournalTitle{Phys. Rev. A}}} \textbf{\bibinfo{volume}{59}},
  \bibinfo{pages}{1829--1834} (\bibinfo{year}{1999}).

\bibitem{Bennett1993}
\bibinfo{author}{Bennett, C.~H.} \emph{et~al.}
\newblock \bibinfo{journal}{\bibinfo{title}{Teleporting an unknown quantum
  state via dual classical and einstein-podolsky-rosen channels}}.
\newblock {\emph{\JournalTitle{Phys. Rev. Lett.}}}
  \textbf{\bibinfo{volume}{70}}, \bibinfo{pages}{1895} (\bibinfo{year}{1993}).

\bibitem{Bennett1992c}
\bibinfo{author}{Bennett, C.~H.} \& \bibinfo{author}{Wiesner, S.~J.}
\newblock \bibinfo{journal}{\bibinfo{title}{Communication via one- and
  two-particle operators on einstein-podolsky-rosen states}}.
\newblock {\emph{\JournalTitle{Phys. Rev. Lett.}}}
  \textbf{\bibinfo{volume}{69}}, \bibinfo{pages}{2881} (\bibinfo{year}{1992}).

\bibitem{Louisell1961}
\bibinfo{author}{Louisell, W.~H.}, \bibinfo{author}{Yariv, A.} \&
  \bibinfo{author}{Siegman, A.~E.}
\newblock \bibinfo{journal}{\bibinfo{title}{Quantum fluctuations and noise in
  parametric processes. i.}}
\newblock {\emph{\JournalTitle{Phys. Rev.}}} \textbf{\bibinfo{volume}{124}},
  \bibinfo{pages}{1646--1654} (\bibinfo{year}{1961}).

\bibitem{Burnham1970}
\bibinfo{author}{Burnham, D.~C.} \& \bibinfo{author}{Weinberg, D.~L.}
\newblock \bibinfo{journal}{\bibinfo{title}{Observation of simultaneity in
  parametric production of optical photon pairs}}.
\newblock {\emph{\JournalTitle{Phys. Rev. Lett.}}}
  \textbf{\bibinfo{volume}{25}}, \bibinfo{pages}{84--87}
  (\bibinfo{year}{1970}).

\bibitem{Fasel2004}
\bibinfo{author}{Fasel, S.} \emph{et~al.}
\newblock \bibinfo{journal}{\bibinfo{title}{High-quality asynchronous heralded
  single-photon source at telecom wavelength}}.
\newblock {\emph{\JournalTitle{New J. Phys.}}} \textbf{\bibinfo{volume}{6}},
  \bibinfo{pages}{163} (\bibinfo{year}{2004}).

\bibitem{Bock2016}
\bibinfo{author}{Bock, M.}, \bibinfo{author}{Lenhard, A.},
  \bibinfo{author}{Chunnilall, C.} \& \bibinfo{author}{Becher, C.}
\newblock \bibinfo{journal}{\bibinfo{title}{Highly efficient heralded
  single-photon source for telecom wavelengths based on a ppln waveguide}}.
\newblock {\emph{\JournalTitle{Opt. Express}}} \textbf{\bibinfo{volume}{24}},
  \bibinfo{pages}{23992--24001} (\bibinfo{year}{2016}).

\bibitem{Pomarico2012}
\bibinfo{author}{Pomarico, E.}, \bibinfo{author}{Sanguinetti, B.},
  \bibinfo{author}{Guerreiro, T.}, \bibinfo{author}{Thew, R.} \&
  \bibinfo{author}{Zbinden, H.}
\newblock \bibinfo{journal}{\bibinfo{title}{{MHz} rate and efficient
  synchronous heralding of single photons at telecom wavelengths}}.
\newblock {\emph{\JournalTitle{Opt. Express}}} \textbf{\bibinfo{volume}{20}},
  \bibinfo{pages}{23846} (\bibinfo{year}{2012}).

\bibitem{Ramelow2013}
\bibinfo{author}{{Ramelow}, S.} \emph{et~al.}
\newblock \bibinfo{journal}{\bibinfo{title}{{Highly efficient heralding of
  entangled single photons}}}.
\newblock {\emph{\JournalTitle{Opt. Express}}} \textbf{\bibinfo{volume}{21}},
  \bibinfo{pages}{6707} (\bibinfo{year}{2013}).

\bibitem{Kaneda2016}
\bibinfo{author}{Kaneda, F.}, \bibinfo{author}{Garay-Palmett, K.},
  \bibinfo{author}{U'Ren, A.~B.} \& \bibinfo{author}{Kwiat, P.~G.}
\newblock \bibinfo{journal}{\bibinfo{title}{Heralded single-photon source
  utilizing highly nondegenerate, spectrally factorable spontaneous parametric
  downconversion}}.
\newblock {\emph{\JournalTitle{Opt. Express}}} \textbf{\bibinfo{volume}{24}},
  \bibinfo{pages}{10733--10747} (\bibinfo{year}{2016}).

\bibitem{Mattle1996}
\bibinfo{author}{Mattle, K.}, \bibinfo{author}{Weinfurter, H.},
  \bibinfo{author}{Kwiat, P.~G.} \& \bibinfo{author}{Zeilinger, A.}
\newblock \bibinfo{journal}{\bibinfo{title}{Dense coding in experimental
  quantum communication}}.
\newblock {\emph{\JournalTitle{Phys. Rev. Lett.}}}
  \textbf{\bibinfo{volume}{76}}, \bibinfo{pages}{4656--4659}
  (\bibinfo{year}{1996}).

\bibitem{Bouwmeester1997}
\bibinfo{author}{Bouwmeester, D.} \emph{et~al.}
\newblock \bibinfo{journal}{\bibinfo{title}{Experimental quantum
  teleportation}}.
\newblock {\emph{\JournalTitle{Nature (London)}}}
  \textbf{\bibinfo{volume}{390}}, \bibinfo{pages}{575--579}
  (\bibinfo{year}{1997}).

\bibitem{Pan1998}
\bibinfo{author}{Pan, J.-W.}, \bibinfo{author}{Bouwmeester, D.},
  \bibinfo{author}{Weinfurter, H.} \& \bibinfo{author}{Zeilinger, A.}
\newblock \bibinfo{journal}{\bibinfo{title}{Experimental entanglement swapping:
  Entangling photons that never interacted}}.
\newblock {\emph{\JournalTitle{Phys. Rev. Lett.}}}
  \textbf{\bibinfo{volume}{80}}, \bibinfo{pages}{3891--2894}
  (\bibinfo{year}{1998}).

\bibitem{Jennewein2000}
\bibinfo{author}{Jennewein, T.}, \bibinfo{author}{Simon, C.},
  \bibinfo{author}{Weihs, G.}, \bibinfo{author}{Weinfurter, H.} \&
  \bibinfo{author}{Zeilinger, A.}
\newblock \bibinfo{journal}{\bibinfo{title}{Quantum cryptography with entangled
  photons}}.
\newblock {\emph{\JournalTitle{Phys. Rev. Lett.}}}
  \textbf{\bibinfo{volume}{84}}, \bibinfo{pages}{4729--4732}
  (\bibinfo{year}{2000}).

\bibitem{Tittel2001}
\bibinfo{author}{Tittel, W.}, \bibinfo{author}{Zbinden, H.} \&
  \bibinfo{author}{Gisin, N.}
\newblock \bibinfo{journal}{\bibinfo{title}{Experimental demonstration of
  quantum secret sharing}}.
\newblock {\emph{\JournalTitle{Phys. Rev. A}}} \textbf{\bibinfo{volume}{63}},
  \bibinfo{pages}{042301} (\bibinfo{year}{2001}).

\bibitem{Sedziak2017}
\bibinfo{author}{Sedziak, K.}, \bibinfo{author}{Lasota, M.} \&
  \bibinfo{author}{Kolenderski, P.}
\newblock \bibinfo{journal}{\bibinfo{title}{Reducing detection noise of a
  photon pair in a dispersive medium by controlling its spectral
  entanglement}}.
\newblock {\emph{\JournalTitle{Optica}}} \textbf{\bibinfo{volume}{4}},
  \bibinfo{pages}{84} (\bibinfo{year}{2017}).

\bibitem{Lasota2018}
\bibinfo{author}{Lasota, M.} \& \bibinfo{author}{Kolenderski, P.}
\newblock \bibinfo{journal}{\bibinfo{title}{Quantum communication improved by
  spectral entanglement and supplementary chromatic dispersion}}.
\newblock {\emph{\JournalTitle{Phys. Rev. A}}} \textbf{\bibinfo{volume}{98}},
  \bibinfo{pages}{062310} (\bibinfo{year}{2018}).

\bibitem{Sedziak2019}
\bibinfo{author}{Sedziak-Kacprowicz, K.}, \bibinfo{author}{Lasota, M.} \&
  \bibinfo{author}{Kolenderski, P.}
\newblock \bibinfo{journal}{\bibinfo{title}{Remote temporal wavepacket
  narrowing}}.
\newblock {\emph{\JournalTitle{Sci. Rep.}}} \textbf{\bibinfo{volume}{9}},
  \bibinfo{pages}{3111} (\bibinfo{year}{2019}).

\bibitem{RPPhotEnc}
\bibinfo{note}{The recalculation can be easily done \emph{e.g.} using the
  script placed on the website
  www.rp-photonics.com/group\_velocity\_dispersion.html.}

\bibitem{Corning}
\bibinfo{note}{See \emph{e.g.} Corning SMF-28e+ Optical Fiber specifications.
  www.corning.com/media/worldwide/coc/documents/\\Fiber/PI1463\_07-14\_English.pdf}.

\bibitem{Shcheslavskiy2016}
\bibinfo{author}{Shcheslavskiy, V.} \emph{et~al.}
\newblock \bibinfo{journal}{\bibinfo{title}{Ultrafast time measurements by
  time-correlated single photon counting coupled with superconducting single
  photon detector}}.
\newblock {\emph{\JournalTitle{Rev. Sci. Instrum.}}}
  \textbf{\bibinfo{volume}{87}}, \bibinfo{pages}{053117}
  (\bibinfo{year}{2016}).

\bibitem{Yan2012}
\bibinfo{author}{Yan, Z.} \emph{et~al.}
\newblock \bibinfo{journal}{\bibinfo{title}{An ultra low noise telecom
  wavelength free running single photon detector using negative feedback
  avalanche diode}}.
\newblock {\emph{\JournalTitle{Rev. Sci. Instrum.}}}
  \textbf{\bibinfo{volume}{83}}, \bibinfo{pages}{073105}
  (\bibinfo{year}{2012}).

\bibitem{Tosi2009}
\bibinfo{author}{Tosi, A.} \emph{et~al.}
\newblock \bibinfo{title}{Ingaas/inp single-photon avalanche diodes show low
  dark counts and require moderate cooling}.
\newblock In \emph{\bibinfo{booktitle}{Society of Photo-Optical Instrumentation
  Engineers (SPIE) Conference Series}}, vol. \bibinfo{volume}{7222},
  \bibinfo{pages}{72221G} (\bibinfo{publisher}{Proc. of SPIE},
  \bibinfo{year}{2009}).

\bibitem{Divochiy2018}
\bibinfo{author}{Divochiy, A.} \emph{et~al.}
\newblock \bibinfo{journal}{\bibinfo{title}{Single photon detection system for
  visible and infrared spectrum range}}.
\newblock {\emph{\JournalTitle{Opt. Lett.}}} \textbf{\bibinfo{volume}{43}},
  \bibinfo{pages}{6085--6088} (\bibinfo{year}{2018}).

\bibitem{Gajewski2016}
\bibinfo{author}{Gajewski, A.} \& \bibinfo{author}{Kolenderski, P.}
\newblock \bibinfo{journal}{\bibinfo{title}{Spectral correlation control in
  down-converted photon pairs}}.
\newblock {\emph{\JournalTitle{Phys. Rev. A}}} \textbf{\bibinfo{volume}{94}},
  \bibinfo{pages}{013838} (\bibinfo{year}{2016}).

\bibitem{Kolenderski2009}
\bibinfo{author}{Kolenderski, P.}, \bibinfo{author}{Wasilewski, W.} \&
  \bibinfo{author}{Banaszek, K.}
\newblock \bibinfo{journal}{\bibinfo{title}{Modelling and optimization of
  photon pair sources based on spontaneous parametric down-conversion}}.
\newblock {\emph{\JournalTitle{Phys. Rev. A}}} \textbf{\bibinfo{volume}{80}},
  \bibinfo{pages}{013811} (\bibinfo{year}{2009}).

\bibitem{Sun2014}
\bibinfo{author}{Sun, P.}, \bibinfo{author}{Guo, J.}, \bibinfo{author}{Wang,
  T.}, \bibinfo{author}{Liu, L.} \& \bibinfo{author}{Feng, J.}
\newblock \bibinfo{journal}{\bibinfo{title}{Method for measuring the intensity
  distribution of a small beam spot with phase retrieval}}.
\newblock {\emph{\JournalTitle{Opt. Laser Eng.}}}
  \textbf{\bibinfo{volume}{57}}, \bibinfo{pages}{109 -- 113}
  (\bibinfo{year}{2014}).

\bibitem{Hwang2003}
\bibinfo{author}{Hwang, W.-Y.}
\newblock \bibinfo{journal}{\bibinfo{title}{Quantum key distribution with high
  loss: Toward global secure communication}}.
\newblock {\emph{\JournalTitle{Phys. Rev. Lett.}}}
  \textbf{\bibinfo{volume}{91}}, \bibinfo{pages}{057901}
  (\bibinfo{year}{2003}).

\bibitem{Korzh2015}
\bibinfo{author}{Korzh, B.} \emph{et~al.}
\newblock \bibinfo{journal}{\bibinfo{title}{Provably secure and practical
  quantum key distribution over 307 km of optical fibre}}.
\newblock {\emph{\JournalTitle{Nature Photonics}}}
  \textbf{\bibinfo{volume}{9}}, \bibinfo{pages}{163--168}
  (\bibinfo{year}{2015}).

\bibitem{Yin2016}
\bibinfo{author}{Yin, H.-L.} \emph{et~al.}
\newblock \bibinfo{journal}{\bibinfo{title}{Measurement-device-independent
  quantum key distribution over a 404 km optical fiber}}.
\newblock {\emph{\JournalTitle{Opt. Express}}} \textbf{\bibinfo{volume}{117}},
  \bibinfo{pages}{190501} (\bibinfo{year}{2016}).

\bibitem{Boaron2018}
\bibinfo{author}{Boaron, A.} \emph{et~al.}
\newblock \bibinfo{journal}{\bibinfo{title}{Secure quantum key distribution
  over 421 km of optical fiber}}.
\newblock {\emph{\JournalTitle{Phys. Rev. Lett.}}}
  \textbf{\bibinfo{volume}{121}}, \bibinfo{pages}{190502}
  (\bibinfo{year}{2018}).

\bibitem{Wang2007}
\bibinfo{author}{Wang, Q.}, \bibinfo{author}{Wang, X.-B.} \&
  \bibinfo{author}{Guo, G.-C.}
\newblock \bibinfo{journal}{\bibinfo{title}{Practical decoy-state method in
  quantum key distribution with a heralded single-photon source}}.
\newblock {\emph{\JournalTitle{Phys. Rev. A}}} \textbf{\bibinfo{volume}{75}},
  \bibinfo{pages}{012312} (\bibinfo{year}{2007}).

\bibitem{Wang2013a}
\bibinfo{author}{Wang, Q.} \& \bibinfo{author}{Wang, X.-B.}
\newblock \bibinfo{journal}{\bibinfo{title}{Efficient implementation of the
  decoy-state measurement-device-independent quantum key distribution with
  heralded single-photon sources}}.
\newblock {\emph{\JournalTitle{Phys. Rev. A}}} \textbf{\bibinfo{volume}{88}},
  \bibinfo{pages}{052332} (\bibinfo{year}{2013}).

\bibitem{Zhu2017}
\bibinfo{author}{Zhu, J.-R.}, \bibinfo{author}{Li, J.}, \bibinfo{author}{Zhang,
  C.-M.} \& \bibinfo{author}{Wang, Q.}
\newblock \bibinfo{journal}{\bibinfo{title}{Parameter optimization in biased
  decoy-state quantum key distribution with both source errors and statistical
  fluctuations}}.
\newblock {\emph{\JournalTitle{Quantum Inf. Process.}}}
  \textbf{\bibinfo{volume}{16}}, \bibinfo{pages}{238} (\bibinfo{year}{2017}).

\bibitem{Eraerds2010}
\bibinfo{author}{Eraerds, P.}, \bibinfo{author}{Walenta, N.},
  \bibinfo{author}{Legre, M.}, \bibinfo{author}{Gisin, N.} \&
  \bibinfo{author}{Zbinden, H.}
\newblock \bibinfo{journal}{\bibinfo{title}{Quantum key distribution and 1
  gbit/s data encryption over a single fibre}}.
\newblock {\emph{\JournalTitle{New J. Phys.}}} \textbf{\bibinfo{volume}{12}},
  \bibinfo{pages}{063027} (\bibinfo{year}{2010}).

\bibitem{Elkouss2013}
\bibinfo{author}{Elkouss, D.}, \bibinfo{author}{Mart\'{i}nez-Mateo, J.},
  \bibinfo{author}{Ciurana, A.} \& \bibinfo{author}{Mart\'{i}n, V.}
\newblock \bibinfo{journal}{\bibinfo{title}{Secure optical networks based on
  quantum key distribution and weakly trusted repeaters}}.
\newblock {\emph{\JournalTitle{IEEE J. Opt. Commun. Netw.}}}
  \textbf{\bibinfo{volume}{5}}, \bibinfo{pages}{316--328}
  (\bibinfo{year}{2013}).

\bibitem{Ciurana2014}
\bibinfo{author}{Ciurana, A.} \emph{et~al.}
\newblock \bibinfo{journal}{\bibinfo{title}{Quantum metropolitan optical
  network based on wavelength division multiplexing}}.
\newblock {\emph{\JournalTitle{Opt. Express}}} \textbf{\bibinfo{volume}{22}},
  \bibinfo{pages}{1576--1593} (\bibinfo{year}{2014}).

\bibitem{Lutz2014}
\bibinfo{author}{Lutz, T.}, \bibinfo{author}{Kolenderski, P.} \&
  \bibinfo{author}{Jennewein, T.}
\newblock \bibinfo{journal}{\bibinfo{title}{Demonstration of spectral
  correlation control in a source of polarization entangled photon pairs at
  telecom wavelength}}.
\newblock {\emph{\JournalTitle{Opt. Lett.}}} \textbf{\bibinfo{volume}{39}},
  \bibinfo{pages}{1481} (\bibinfo{year}{2014}).

\bibitem{Dmitriev1999}
\bibinfo{author}{Dmitriev, V.~G.}, \bibinfo{author}{Gurzadyan, G.~G.} \&
  \bibinfo{author}{Nikogosyan, D.~N.}
\newblock \emph{\bibinfo{title}{Handbook of Nonlinear Optical Crystals}}
  (\bibinfo{publisher}{Springer}, \bibinfo{year}{1999}), \bibinfo{edition}{3rd}
  edn.

\bibitem{Kraus2005}
\bibinfo{author}{Kraus, B.}, \bibinfo{author}{Gisin, N.} \&
  \bibinfo{author}{Renner, R.}
\newblock \bibinfo{journal}{\bibinfo{title}{Lower and upper bounds on the
  secret-key rate for quantum key distribution protocols using one-way
  classical communication}}.
\newblock {\emph{\JournalTitle{Phys. Rev. Lett.}}}
  \textbf{\bibinfo{volume}{95}}, \bibinfo{pages}{080501}
  (\bibinfo{year}{2005}).

\end{thebibliography}

\section*{Acknowledgements}

The authors acknowledge financial support by the Foundation for Polish Science (FNP) (project First Team co-financed by the European Union under the European Regional Development Fund), Ministy of Science and higher Education, Poland (MNiSW) (grant no.~6576/IA/SP/2016) and National Science Centre, Poland (NCN) (Sonata 12 grant no.~2016/23/D/ST2/02064). We wish to thank National Laboratory of Atomic, Molecular and Optical Physics, Torun, Poland for the support.

\section*{Author contributions statement}

M.L. performed most of the analytical and numerical calculations. P.K. supervised the research. Both authors contributed to writing the paper.

\section*{Additional information}

\textbf{Competing interests:} The authors declare no competing interests.

\end{document}